\documentstyle[12pt]{article}
\begin{document}
\vspace{-2.6cm}

\title{\large\bfseries
PHYSICAL MODEL OF LEPTONS: MASSIVE ELECTRONS, \\
MUONS, TAUONS AND THEIR MASSLESS NEUTRINOUS }
\vspace{-2.5cm}

\author{ Josiph Mladenov Rangelov\,,\\
Institute of Solid State Physics, Bulgarian Academy of Sciences,\\
72\,Tsarigradsko chaussee\,,\,1784\,\,Sofia\,,\,Bulgaria\,.}
\date{}
\vspace{-2cm}

\maketitle

\begin{abstract}
 The physical model (PhsMdl) of the leptons is offered by means of the
PhsMdls of the vacuum and electron,described in our recent works. It is
assumed that the vacuum is consistent by dynamides, streamlined in junctions
of some tight crystalline lattice. Every dynamide contains a neutral pair of
massless point-like (PntLk) contrary elementary electric charges (ElmElcChrgs):
electrino $(-)$ and positrino $(+)$. The PntLk ElmElcChrgs of the massless
electrino and positrino of some dynamide in the fluctuated vacuum may been
excited or deviated by means of some energy, introduced by some photon or
other micro particles (MicrPrts). The massless leptons (neutrinos) are
neutral long-living solitary spherical vortical oscillation excitations of
the uncharged fluctuating vacuum. The massive leptons are charged long-living
solitary spherical vortical excitations of its fine spread (FnSpr) elementary
electric charge (ElmElcChrg). So-called zitterbewegung is self-consistent
strong-correlated vortical harmonic oscillation motion of the FnSpr ElmElcChrg
of massive leptons. Different leptons have different self-consistent
strong-correlated vortical harmonic oscillation motion of different sizes of
their FnSpr ElmElcChrg, which is determined by their Kompton length $\lambda\,
=\frac{h}{m\,C} $, where $m$ is the mass of the massive leptons. At mutual
transition of one massive lepton into another massive lepton its PntLk
ElmElcChrg move up by dint of weak interaction in the form of the charged
intermediate vector meson $W$ from one self-consistent strong-correlated
vortical harmonic oscillation motion of one size into another self-consistent
strong-correlated vortical harmonic oscillation motion of another size.
\end{abstract}

 Although up to the present nobody of scientists distinctly knows are there
some elementary micro particles (ElmMicrPrts) as a stone of the micro world
and what the elementary micro particle (ElmMicrPrt) means, there exists an
essential possibility for clear and obvious scientific consideration of the
uncommon quantum behaviour and unusual relativistic dynamical parameters of
all the relativistic quantized micro particle (QntMicrPrt) by means of our
transparent surveyed PhsMdl. It is well known that the physical model (PhsMdl)
presents at us as an actual ingradient of every good physical theory (PhsThr).
It may be used as for an obvious visual teaching the unknown occurred physical
processes within the investigated phenomena, It turned out that all leptons
are elementary micro particles (ElmMicrPrt) of two kinds: a charged massive
and uncharged massless. The massless leptons (neutrinos) are neutral
long-living solitary isotropic three dimensional relativistic quantized
(SltIstThrDmnRltQnt) spherical vortex harmonic oscillation (SphrVrtNrmOsc)
excitations of the neutral fluctuated vacuum (FlcVcm) without a self-energy
at a rest. The massive leptons are charged long-living solitary isotropic
three dimensional relativistic quantized (SltIstThrDmnRltQnt) spherical
vortical harmonic oscillation (SphrVrtHrmOsc) motion of its fine spread
(FnSpr) elementary electric charge (ElmElcChrg). The PhsMdl of the charged
massive lepton will be presented by the PhsMdl of one of them, the DrEl.
Therefore so-called zitterbewegung by Schrodinger,which is self-consistent
strong-correlated vortical harmonic oscillation (VrtHrmOscs) motion of the
FnSpr ElmElcChrg of one of massive leptons, electron, is generalized for all
charged massive leptons. Therefore different charged massive leptons have
different self-consistent strong-correlated vortical harmonic oscillation
motion of different sizes of their FnSpr ElmElcChrg, which is determined by
their Kompton length $\lambda\,= \frac{h}{m\,C}$, where $m$ is the mass of
the massive leptons. At mutual transition of one massive lepton into another
massive lepton its PntLk ElmElcChrg move up by means of weak interaction in
the form of the charged intermediate vector meson $W$ from one self-consistent
strong-correlated (VrtHrmOscs) motion of one size into another self-consistent
strong-correlated (VrtHrmOscs) motion of another size.

 The PhsMdl of the DrEl have offered in all my work in resent nineteen years
for bring of light to the physical interpretation of physical cause of the
uncommon quantum behaviour of the Schrodinger electron (SchEl) and the
relativistic behaviour of the DrEl and give the thru physical interpretation
and sense of all its dynamical parameters. Our PhsMdl of the DrEl explain as
the physical causes for its unusual stochastic classical dual wave-corpuscular
behaviour and so give a new cleared picturesque physical interpretation with
mother wit of the physical means of its relativistic dynamical parameters. In
our transparent surveyed PhsMdl of the DrEl one will be regarded as some point
like (PntLk) ElmElcChrg, taking simultaneously part in four different motions:

 A) The isotropic three-dimensional relativistic quantized (IstThrDmnRltQnt)
Einstein's stochastic (EinStch) boson harmonic shudders(BsnHrmShdrs) as a
result of momentum recoils (impulse kicks), forcing the charged QntMicrPrt at
its continuously stochastical emissions and absorptions of own high energy
(HghEnr) virtual photons (StchVrtPhtns) by its PntLk ElmElcChrg. This jerky
motion display almost Brownian classical stochastic behaviour (BrnClsStchBhv)
with a light velocity C during a small time interval $\tau_1$, much less then
the period $T$ of the IstThrDmnRltQnt Schrodinger fermion (SchFrm) vortical
harmonic oscillation (VrtHrmOsc) motion and more larger then the time
interval $\tau_o$ of the stochastically emission or absorption of the Hgh-Enr
StchVrtPhtn by its PntLk ElmElcChrg. The display of the IstThrDmnRltQnt
EinStch BsnHrmShds can be observed at the light (RlPhtn) scattering of free
DrEl. Indeed, in a consequence of the above investigation we may really
consider the IstThrDmnRltQnt FrthStch VrtHrmOscMtn's "trajectory" of the DrEl
turns into cylindrically fine spread path, which has a form of the distorted
figure of eight. The participate the FnSpr ElmElcChrg of DrEl in
IstThrDmnRltQnt FrthStch VrtHrmOscMtn well spread (WllSpr) the ElmElcChrg of
the SchEl. Further the behaviour of the DrEl's FnSpr ElmElcChrg may be
treated as a nearly Brownian classical stochastic behaviour (BrnClsStchBhv)
of the ClsMicrPrt, taking part in the relativistic random trembling motion
(RltRndTrmMtn), during the time intervals $\tau_1$, much less then the period
$T$ of the SchFrm VrtHrmOscs and more larger then the time interval $\tau$ of
the emission and absorption of the Hgh-Enr StchVrtPhtn, the QntElcMgnFld of
which form the RslSlfCnsVls of own RslQntElcMgmFld of the DrEl's FnSpr
ElmElcChrg.  In a consequence of this the product  $2\,C\,\tau$ may be
considered as the space parth values of the DrEl's PntLk ElmElcChrg, which is
passed in the time interval $2\,\tau$ of its absorption and emission (or
scattering) of some RlPhtn. In a consequence of this we can easily understand
by our felicitous PhsMdl why the classical radius of the LrEl entirely
coincides with the size $\sqrt{\langle\,\xi^2\, \rangle}$ of a spherical fine
spread spot with an effective scattering surface $\pi\,\langle\xi^2\rangle$.
Therefore the IstThrDmnRltQnt SchFrm VrtHrmOscMtn's trajectory turns into fine
spread path of a cylindrical shape with different radii.  Therefore the size
of this smallest IstThrDmnRltQnt EinStch BznHrmShd's motion of the DrEl's
PntLk ElmElcChrg could be determined by the Thompson total cross section.
Indeed the averaged cross section of the cylindrical spread path of the
IstThrDmnRltQnt StchBsnHrmOscMtn can be determined by the effective total
cross section of Thompson of RlPhtn's scattering from the free DrEl's, which
has a space distribution with a spherical symmetry may be easily obtained by
means of the following simple relations, well-known by us from the classical
StchMch :
\begin{equation}\label{l}
\sigma\,=\,\pi\,\langle\,\xi^2\,\rangle\,=
\,\frac{3\pi}{2}\,\left[\,\frac{4\,e^2}{3mC^2}\,\right]^2\,=
\,\frac{8\pi}{3}\,\left[\,\frac{e^2}{mC^2}\,\right]^2\,
\end{equation}

It is very important to make here, that this EinStchBznHrmShds have the
smallest size and the fastest velocity and therefore the described smallest
cross section may be represented as a described roughly by the space
distribution of the DrEl's FnSpt ElmElcChrg :
\begin{equation}\label{z3}
\,|\,\Upsilon_o\,(\varrho)\,|^2\,=\,\left[(\frac{3}{5\sqrt{\pi}})^3
\,(\frac{mC^2}{e^2})^3\right]\,\exp(\frac{-\varrho^2}{\kappa_o^2})
\end{equation}

where $\kappa_o$ is parameter of the EinStch BsnHrmShds $(\kappa_o\,=\,(5/3)\,
\frac{e^2}{mc^2}\,$. The space distribution (\ref{z3}) of the DrEl's FnSpt
ElmElcChrg is described roughly by an OrbWvFnc $\Upsilon$, having the
following form :
\begin{equation}\label{z4}
\,\Upsilon_o(\varrho)\,=\,(\sqrt{\pi}\kappa_o)^-\frac{3}{2}\,
\exp(\frac{-\varrho^2}{2\kappa_o^2})
\end{equation}

B) The isotropic three-dimensional relativistic quantized (IstThrDmnRltQnt)
Schrodinger fermion vortical harmonic oscillation motion (SchFrmVrtHrmOscMtn).
In a consequence of such jerks of the PntLk ElmElcChrg along the
IstThrDmnRltQnt EinStchHrmOscShdMtn "trajectory" the "trajectory" of
the DrEl's FnSpt ElmElcChrg, participating in the IstThrDmnRltQnt
SchFrm VrtHrmOscMtn takes a strongly broken shape. Only after the
correspondent averaging over the "trajectory" of the IstThrDmnRltQnt
EinStch BznHrmShdMtn we may obtain the fine spread "trajectory" of the
IstThrDmnRltQnt SchFrm VrtHrmOscs' one, having got the form of the distorted
figure of an eight. Only such a motion along a spread uncommon "trajectory"
of the DrEl's FnSpt ElmElcChrg could through a new light over the SchEl's
well spread (WllSpr) ElmElcChrg's space distribution and over the spherical
symmetry of the SchEl's WllSpr ElmElcChrg. It turns up that all relativistic
dynamical properties of the DrEl are results of the participation of its fine
spread (FnSpr) ElmElcChrg in the Schrodinger's self-consistent fermion
strongly correlated harmonic oscillations motion. This self-consistent
strongly correlated IstThrDmnRltQnt SchFrm VrtHrmOsc's motion may be described
mathematically correctly by means of the four components of its total wave
function (TtlWvFnc) $\Psi$ and four Dirac's matrices\,;\,$\alpha_j\,
(\gamma_j)$ and $\beta\,(\gamma_o)$.

 It turned out that all the massive leptons are sums of the corresponding
massless lepton (neutrino) and the FnSpr ElmElcChrg, which participates in
some kind of a badly known but powerfully correlated self-consistent fermion
motion, called zitterbewegung. The different massive leptons are distinguished
between them-self by the amplitude size and frequency of its IstThrDmnRltQnt
SchFrm VrtHrmOscs, called zitterbewegung. Hence the different aroma of the
leptons are different self-consistent spherical excitations of its fine
spread (FnSpr) elementary electric charge (ElmElcChrg), which participates in
isotropic three dimensional relativistic quantized (IstThrDmnRltQnt)
Schrodinger's fermion (SchrFrm) vortical harmonic oscillations (VrtHrmOscs)
of different sizes and frequency, which are determined by their Kompton
length $\lambda\,=\,\frac{h} {m\,C} $,where $m$ is the mass of the massive
leptons, at different energies, which are determined by their mass $m$.

 In such a way the deviated FnSpr ElmElcChrg can creates own self-consistent
resultant quantized electromagnetic fields (QntElcMgnFlds) by dint of own
high energy stochastic virtual photons (StchVrtPhtns), emitted by itself at
different points of its zitterbewegung trajectory and in different moments in
positions of the self-consistent powerfully correlated SchrFrm VrtHrmOscs
motion.  Therefore the rest self-energy of the ElmMicrPrt $E_o\,=\,mc^2$ is
created in a results of the electromagnetic interaction (ElcMgnIntAct) between
its PntLk ElmElcChrg and MgnDplMm with the electric intensity (ElcInt) and
magnetic intensity (MgnInt) of their own QntElcMgnFld. The own resultant
QntElcMgnFld of the PntLk ElmElcChrg of the charged ElmMicrPrt is a result of
the sum of the QntElcMgnFlds of constantly emitted stochastic VrtPhtns by same
PntLk ElmElcChrg from its different positions of the space in the zitterbewegung
trajectory at different moments of a time at its self-consistently powerfully
correlated SchrFrm VrtHrmOsc motion. Therefore the different ElmMicrPrts,
which are different aromas of the leptons, may be considered as stable
excitations of different energy in the uncharged fluctuating vacuum (FlcVcm).

 C)The isotropic three-dimensional nonrelativistic quantized (IstThrDmnNrlQnt)
Furthian stochastic boson (FrthStchBsn) circular harmonic oscillation motion
(CrcHrmOscMtn) of the SchEl as a result of the permanent ElcIntAct of the
electric intensity (ElcInt) of the resultant QntElcMgnFld of all the low energetic (LwEmr) StchVrtPhtns, existing within the FlcVcm and
generated by dint of the VrtPhtn's stochastic exchange between them. The
SchEl's motion and its unusual quantized behaviour, described in the
NrlQntMch may be easily understood by assuming it as a forced random
trembling motion (RndTrmMtn) upon a stochastic joggle influence of the
StchVrtPhtns scattering from some FrthQntPrt. Therefore the RndTrmMtn can be
approximately described through some determining calculations by means of
both the laws of the Maxwell ClsElcDnm and the probabilistic laws of the
classical stochastic theory (ClsStchThr). But in a principle the exact
description of the SchEl's uncommon behaviour can be carry into a practice by
means only of the laws of NrlQntMch and ClsElcDnm.

 Since then it is easily to understand by means of upper account that if the
ClsMicrPrt's motion is going along the clear definitived smooth thin trajectory
in accordance with the NrlClsMch, then the QntMicrPrt's motion is perform in
the form of a roughly cylindrical spread path of a cylindrical shape with
different radii with centers on a strongly broken line with quite unordered
in its direction small straight lines of the RndTrbMtn near the classical one
of any NtnClsPrt within the NrlClsMch. As a result of that we can suppose
that the unusual dualistic behaviour of QntMicrPrt can be described by dint of
\begin{equation}\label{t}
\,r_j\,=\,\bar {r}_j\,+\,\delta {r}_j\quad; \quad
p_j\,=\,\bar {p}_j\,+\,\delta {p}_j\quad;
\end{equation}

 It turns up that all the quantized dynamical properties of the SchEl are
results of the participation of its WllSpt ElmElcChrg in the isotropic three
dimensional nonrelativistic quantized (IstThrDmnNrlQnt) Furth's stochastic
(FrthStch) boson harmonic oscillations(BsnHrmOscs). It is used as for a visual
teaching the occurred physical processes within the investigated phenomena, so
for doing them equal with the capacity of its mathematical correct description
by the mathematical apparatus of the both the quantum mechanics : the
nonrelativistic (NrlQntMch) and relativistic (RltQntMch).

 D) The classical motion of the LrEl along an well contoured smooth and thin
trajectory realized in a consequence of some classical interaction (ClsIntAct)
of its over spread (OvrSpr) ElmElcChrg, bare mass or magnetic dipole moment
(MgnDplMm) with some external classical fields (ClsFlds), described by well
known laws of the Newton nonrelativistic classical mechanics (NrlClsMch). This
motion may be finically described by virtue of the laws of both the NrlClsMch
and the classical electrodynamics (ClsElcDnm);

 We must draw attention here that two massive leptons of same aroma may been
distinguished also by direction of its twirl. But if only massless leptons
have primary twirl the massive leptons may have both twirls. We must draw
attention also here that the neutral spherical vortex excitations may been
distinguished only by direction of its twirl. As there are possibility for
two opportunity twirls then neutrino has left-handed twirl and therefore the
spin direction of the neutrino is antiparallel of its impulses direction while
the antineutrino has right-handed twirl and therefore the spin direction of
the antineutrino is parallel of its impulses direction. For knowing this it is
very interesting  why many theoretical physicists debate frequently why there
are as left-handed photons so there are and right-handed photons, but nobody
speaks about photons and antiphotons. However as because massless leptons
(neurtinos) participate in the weak interactions therefore the primary twirl
may been observanted in the weak interaction. Therefore many of them assert
that there is asymmetry between weak interaction and electromagnetic
interaction. Indeed if somebody of them call the left-handed photon a photon
and call the right-handed photon an antiphoton, then both interactions, weak
and electromagnetic should been semantically.

 In is easily to perceive that at the attentive analysis of the decay formulas
of the different lepton aromas :
\begin{equation}\label{a}
\tau^{+} \longrightarrow \mu^{+} + \tilde{\nu}_\tau\,+\,\nu_\mu \quad,
\tau^{+} \longrightarrow e^{+} + \tilde{\nu}_\tau\,+\,\nu_e \quad,
\mu^{+} \longrightarrow e^{+} + \tilde{\nu}_\mu\,+\,\nu_e \quad,\\
\end{equation}

\vspace{-0.7cm}

\begin{equation}\label{b}
\tau^{-} \longrightarrow \mu^{-} + \nu_\tau + \tilde{\nu}_\mu\quad,
\tau^{-} \longrightarrow e^{-} + \nu_\tau + \tilde{\nu}_e \quad,
\mu^{-} \longrightarrow e^{-} + \nu_\mu + \tilde{\nu}_e \quad,\\
\end{equation}

It is seen by means of these formulas that we are ability to write the
follow equations:
\begin{equation}\label{c}
\tau^{+} = W^{+} + \tilde{\nu}_\tau ,\quad \mu^{+} = W^{+} + \tilde{\nu}_\mu ,
\quad e^{+} = W^{+} + \tilde{\nu}_e ,  \\
\end{equation}

\vspace{-0.7cm}

\begin{equation}\label{d}
\tau^{-} = W^{-} + \nu_\tau ,\quad \mu^{-} = W^{-} + \nu_\mu ,
\quad e^{-} = W^{-} + \nu_e , \\
\end{equation}

 It is easily to understand from upper that from both group decay formulas we
are ability to assume the existence of the follow decay reactions:
\begin{equation}\label{f}
W^{+} \longrightarrow \tau^{+} + \nu_\tau , \quad W^{+} \longrightarrow
\mu^{+} + \nu_\mu , \quad W^{+} \longrightarrow e^{+} + \nu_e , \\
\end{equation}

\vspace{-0.7cm}

\begin{equation}\label{g}
W^{-} \longrightarrow \tau^{-} + \tilde{\nu}_\tau , \quad W^{-}
\longrightarrow \mu^{-} + \tilde{\nu}_\mu , \quad W^{-} \longrightarrow
e^{-} + \tilde{\nu}_e , \\
\end{equation}

 The upper decay reaction remain us about the emmiting of a real photon
(RlPht) by stimulated atom. Really  as we well know the Rlpht is no found
within stimulated atom before its radiation.Therefore we can speak only
about the electrino,positrino and neutrinos.That is why we can use the
following formal registration:
\begin{equation}\label{f1}
\tau^{+} = (+) + \tilde{\nu}_\tau , \quad \mu^{+} = (+) + \tilde{\nu}_\mu ,
\quad  e^{+} = (+) + \tilde{\nu}_e , \\
\end{equation}

\vspace{-0.7cm}

\begin{equation}\label{g1}
\tau^{-} = (-) + \nu_\tau  , \quad \mu^{-} = (-) + \nu_\mu , \quad e^{-} =
(-) + \nu_e \\
\end{equation}

But the upper decay reaction have no means that the charged intermediate
vector bosons $W$ are composed from some aroma lepton and its neutrino, as
in reality the charged intermediate vector bosons $W$ are created only in the
time of transfer beginning od the PntLk ElmElcChrg from one ElmPrt to other
ElmPrt and are existed only during the time interval of same transfer. We can
distinctly see this from the following equation:
\begin{equation}\label{i}
\quad Z^o\,=\,\nu_{e} \,+\,\tilde {\nu}_{e} \quad ,
\quad Z^o\,=\,\nu_{\mu}\,+\,\tilde {\nu}_{\mu} \quad ,
\quad Z^o\,=\,\nu_{\tau}\,+\,\tilde {\nu}_{\tau} \quad ,
\end{equation}

 The equation (\ref{i}) shows that the neutral intermediate vector boson $Z^o$
contains two opposite PntLk ElmElcChrgs (electrino and positrino). During the
time interval of the decay of the neutral intermediate vector boson $Z^o$ both
its opposite PntLk ElmElcChrgs reconstruct own self-consistent motions and
annihilate, formating one dynamide, while its flat motions decay in two
parallel neutral spherical vortical excitations: neutrino and antineutrino of
same aroma.

 But the upper decay reaction have no means that the some massive charged
lepton is composed from same aroma neutrino and the PntLk ElmElcChrg of the
negative charged intermediate vector bosons $W^{-}$ and every massive
antilepton is sum of its antineutrino and positive charged intermediate
vectorial boson $W^{+}$. Indeed, although the decay of one lepton from another
lepton is accompanied with leaping its negative PntLk ElmElcChrgthe in a state
of the negative charged intermediate vectorial boson $W^{-}$ from one neutrino
to another neutrino. But this decay don't means that every massive lepton is
sum of its neutrino and negative charged intermediate vectorial boson $W^{-}$
Really we can think that some massive lepton receives energy from fluctuating
vacuum (FlcVcm) by dint of some virtual photon (VrtPhtn) or virtual gluon
(VrtGln) and therefore it generate neutrino of its aroma in time moment of
its transition in state of the negative charged intermediate vectorial boson
$W^{-}$ In such the way the unstable negative charged intermediate vectorial
boson $W^{-}$ gives back of FlcVcm borrowing from it energy in form of the
VrtPhtn or VrtGln and after that its negative PntLk ElmElcChrg emits the
antineutrino of same aroma, which aroma has the lepton massive state, which
it occupy, without disintegrating itself of the negative charged intermediate
vectorial boson $W^{-}$ and massless neutrino.

 However there is possibility to understand by means of upper decay relation
 why exist lepton and antilepton numbers and weak charges, satisfying the
 conservation laws and why absent the electric charge from symmetry law.
 Indeed, if negative charged intermediate vectorial boson $W^{-}$ has spin
 minus $\hbar $, then in order to some lepton with spin a minus half $\hbar
 $, we must add only the antineutrino with spin a half $\hbar $. If we wish
 to  make some antilepton, then to neutrino with a spin minus half $\hbar$
 we must accompanied with leaping the positive charged intermediate
 vectorial boson $W^{+}$ with a spin $\hbar$.

 The upper decay group formulas teach us that the PntLk ElmElcchrg of
different leptons participates in some IstThrDmnRltQnt self-consistent and
powerful correlated SchrFrnHrmOsc motion (zitterbewegung) of different sizes
and at different energies. Therefore at the mutual transitions between them
there give birth of pair neutrino $\nu _l $ and antineutrino $\tilde {\nu_l}
$ of the same aroma and the PntLk ElmElcChrd pass from own neutrino
(antineutrino) to the new birthed neutrino (antineutrino) in the form of the
charged intermediate vectorial boson $W$. This is a natural way from which we
can see the unity of the field neutral excitations in the FlcVcm and its
substantial charged excitations, offered by modernity relativistic quantum
mechanics (RltQntMch), quantum electrodynamics (QntElcDnm) and quantum theory
of field (QntThrFld). The electric interaction (ElcIntAct) of the resultant
QntElcMgnFld of all the StchVrtPhtns, exchanged from the FlcVcm with the
ElmMicrPrt's PntLk ElmElcChrg creates their diverse oscillations along its
classical well contoured smooth and thin trajectory, spread and turned it
into wide path, described by its OrbWvFnc $(\Psi)$ within the nonrelativistic
quantum mechanics (NrlQntMch) .

 It turns out that we describe only three aroma massive and massless leptons.
In approximation of a mathematical correct substitution of the three
one-dimensional powerful correlated fermion harmonic oscillations with
three one-dimensional independent boson harmonic oscillations the size of
each aroma of leptons is determined by length of its Kompton wave $\lambda\,=
\,\frac{h}{m\,C} $. As the masses of three massive lepton aromas have very big
different values $(1, 207, 1785) m_e C^2 $, then and the size of each lepton
aroma must very strong differ one from other. Here I wish to show one very
interesting  coincidence. Indeed, if the total energy of each charged massive
leptons is a sum of its self-energy of rest $ m.C^2$ and of  the potential
energy of its PntLk ElmElcChrg in own averaged QntElcMfnFld $(2/3)\,\frac{e^2}
{C\hbar}\,m.C^2$, then the total energy of the muon $\mu$ $m_{\mu}.C^2$ is
equal of the sum of its self-energy of rest $ m_{\mu}.C^2$ and of the energy
of the electron $m_e.C^2$, which is the potential energy of its PntLk
ElmElcChrg in own averaged QntElcMgnFld $\frac{2}{3}\,\frac{e^2}{C\hbar}
\,m_{\mu}.C^2 $. Indeed :
\begin{equation}\label{k}
\,m_e \,\left\{\,1\,+\,(3/2)\frac{C\hbar}{e^2}\,\right\}\,\cong\,
m_{\mu} \quad {\rm or} \quad \left(\,1\,+\,205.54\,\right)\,\cong\,206.7
\end{equation}

 Really by means of upper equation (\ref{k}) we can assert that the total energy
$m_e .C^2$ of the electron is an equal of the potential energy of the PntLk
ElmElcChrg of muon in own averaged QntElcMgnFld. From research of the upper
decay formulas we can understand that the weak interaction is result of the
charged (or neutral) intermediate vector bosons $W$ (or $Z$) mutually
interchange between leptons and others elementary micro particles (ElmMicrPrt).
The formation of the lepton with a spin of the half of $\hbar$ from charged
intermediate vector boson $W$ with a spin of one $\hbar$ and another lepton
with a spin of the half of $\hbar$ determines the choice rule of the
participating the lepton in this interaction.

 The relation $E^2 = p^2c^2 + m^2c^4$ between the energy, impulse and mass of
the ClsMacrPrt may be obtained through the use of the Maxwell's equations of
the classical electrodynamics (ClsElcDnm),as a result of the relation between
the harmonic oscillations of the impulse of a charged ClsMacrPrt and the
vector-potential of its ClsElcMgnFld. The parameters of own resultant
QntElcMgnFld in the point of the moment positions of the QntMicrPrt's PntLk
ElmElcChrg and value of its rest-self energy may be determined by the agency
of the mathematical apparatus of the RltQntMch, QntThrFld and QntElcDnm. The
created by this way own resultant QntElcMgnFld have zero values of the electric
intensity (ElcInt) of own resultant QntElcFld in the point of the moment position
of a MicrPrt's PntLk ElmElcChrg and doubled value of the magnetic intensity
(MgnInt) of own resultant QntMgnFld in this point in a respect of the MgnInt of
the ClsMgnFld, created by the small spread (SmlSpr) ElmElcChrg, participating
in isotropic three dimensional relativistic classical (IstThrDmnRltCls) Debay
boson harmonic oscillations (DbBsnHrmOscs) with same the energy.

The electric interaction (ElcIntAct) of the resultant QntElcMgnFld of the
StchVrtPhtns from the fluctuating vacuum (FlcVcm) with a ElmMicrPrt's WllSpr
ElmElcChrg creates their diverse oscillations along its classical trajectory,
spread and turned into an wide path within the nonrelativistic quantum mechanics
(NrlQunMch). Such IstThrDmnRltQnt Furthian stochastic boson circular harmonic
oscillation motion (FrthStchBsnCrcHrmOscMtn) secures the existence of an
additional mechanical moment (MchMm) of the QntMicrPrt, as and anomalous part
of its MgnDplMm. The energy of the StchVrtPhtns exchanged between
QntMicrPrt's WllSpr ElmElcChrg and the FlcVcm gives a possibility of the
QntMicrPrts to make tunneling through potential barriers, which are
impassable in a classical way. The QntMicrPrt gathers from the FlcVcm at its
IstThrMrnRltQnt BrnStchBsnCrcHrmOscMtn for potential energies of an averaged
ElcFld of its WllSpr ElmElcChrg. Therefore the potential energies of this
field don't take part in equations between an inserted energy in the
beginning of the birth of the ElmMicrPrts and obtained after its end.

 These isotropic three dimensional relativistic quantized (IstThrDmnRltQnt)
Schrodinger fermion vortical harmonic oscillation (SchFrmVrtHrmOsc) motion
(zitterbewegung) of the ElmMicrPrt's FnSpr ElmElcChrg correspond to its inner
harmonical motion, introduced by Louis de Broglier. The Schrodinger's
zitterbewegung is some powerful correlation fermion self-consistent motion,
who minimizes the rest self-energy of the ElmMicrPrt and secures the
continuous stability of the Schrodinger's wave package (SchWvPck) in the
space inanalogous of the Debay wave package (DbWvPck). In such a way we
understand that the energetical advantage of the self-consistent strong
correlated zitterbewegung, which minimizes the energy of ElcMgnSlfAct between
the FnSpr ElmElcChrg and its own resultant QntElcMgnFld secures the stability
of the SchrWvPck in the space and time.

 The emission and absorption of high energy StchVrtPhtns by the PntLk
ElmElcChrg of the charged ElmMicrPrt forces itself to make a isotropic
three-dimensional relativistic quantized (IstThrDmnRltQnt) Einstein
stochastic (EinStch) boson harmonic oscillation motion (BsnHrmOscMtn), which
makes the smooth trajectory of the SchrFrm VrtHrmOscMtn a thickly and
strongly broken line with shortest and very disordered straight lines. The
fine spread (FnSpr) of the SchrFrm VrtHrmOscMtn's trajectory by very rapid
jerk EinStch BsnHrmOscMtn may be observed at the scattering of the light
(RlPhtns) on the free Dirac electrons (DrEl). Indeed, the IstThrDmnRltQnt
SchrFrm VrtHrmOscMtn's trajectory turns into roughly spread path of a
cylindrical shape with different radii. The averaged cross section of the
cylindrical spread path of the IstThrDmnRltQnt StchBsn VrtHrmOscMtn can be
determined by Thompson total cross section $\sigma\,=\,\frac{8\pi}{3} \,
\left[\,\frac{e^2}{mC^2}\,\right]^2\,$ of the of the light (RlPhtns) at the
free DrEl ,which determines the classical radius $r_o\,=\,\frac{2e^2}{mC^2}
\,\sqrt{2/3}$ of the LrEl.

 Although till now nobody knows what the ElmmICRpRT means, there exists a
possibility for a consideration of the unusual behaviour of the quantized
micro particles (QntMicrPrts), of such as leptons and adrons by dint of an
analogy with the transparent surveyed PhsMdl of the DrEl. By our PhsMdl of
the leptons at transition of the PntLk ElmElkChrg from one massless leptons
(its neutrino) to another massless leptons (its neutrino) it takes form of
a charged intermediate vector meson $W$. In such a way the weak interaction
between two leptons may be realized by a transition of one PntLk ElmElcChrg
from one neutrino (anti neutrino) to another neutrino (antineutrino) in the
form of a charged intermediate vector meson $W$.

 It seems to me the existance of two very interesting facts,having common
physical cause.The first is concurrence of the energy of one degree of
freedom in charged lepton $\mu$-meson and in charged admeson $\pi$-meson.
Indeed, if in isotropic three dimensional solitary vortical harmonic
ocsillations of FnSpr ElmElcChrg of $\mu$-meson have three degrees of freedom
and therefore $3\,\hbar\,\omega = 2\,m\,C^2 = 213.2$Mev. Hence the energy of
one degree of freedom can be determined $\frac{\hbar\,\omega}{2} = 35.5 $Mev.
If we take into consideration that the FnSpr ElmElcChrg of $\pi$-meson takes
participation in two quasi-plane circular harmonic oscillations with
opportunity orientations and therefore has energy 2$\hbar\omega = 139.6$Mev.
Hence the energy of one degree of freedom can be determined $\frac{\hbar\,
\omega}{2} = 34.9 $Mev. As we can see by comparision of two results this
coincidence is very accurate. On this reason we can assume that the areas of
their oscillations must also coincidence and therefore the OrbWvFnc of both
FnSpr ElmElcChrg. May be therefore the decay of the positive (negative)
charged $\pi$-meson in $100\%$ occurs through the positive (negative)
$\mu$-meson and $\mu$-neutrino (antineutrino). This second coincidence gives
us many correct answer of the question for inner structure of the elementary
micro particles (ElmMicrPrts).

 I think that is very interesting to wtite the equations of the
incomprehencible decay down:
\begin{eqnarray}\label{an}
\pi^{+} \Longrightarrow  W^{+} \Longrightarrow  \mu^{+} + \nu_\mu \quad,
\pi^{-} \Longrightarrow  W^{-} \Longrightarrow  \mu^{-} + \tilde{\nu}_\mu ,
\end{eqnarray}

\vspace{-0.7cm}

\begin{eqnarray}\label{cn}
\pi^{+} \Longrightarrow  W^{+} \Longrightarrow  e^{+} + \nu_e \quad,
\pi^{-} \Longrightarrow  W^{-} \Longrightarrow  e^{-} + \tilde{\nu}_e ,
\end{eqnarray}

 As for the radiation of the spontaneous real photon (RlPhtn) from the
excitative atom it is necessary the presence of the virtual photon (VrtPhtn)
for the creation of the electric dipole moment (ElcDplMm), so for the decay
of a charged $\pi$-meson it is necessary the presence of a virtual gluon for
an overturning of the spin of one of its quarks, by which charged $\pi$-meson
turns into charged virtual $\rho$-meson, which can immediately decay into
charged intermediate vector boson $W$. At the subsequent transfer of the
charged intermediate vector bozon $ W $ in a pair of massive and massless
leptons of equal aroma the participating in the decay gluon go back in the
FlcVcm. Therefore instead of upper decays we must used the following equations :
\begin{eqnarray}\label{gn}
\pi^{+} + \delta \Longrightarrow W^{+} \Longrightarrow \mu^{+} + \nu_\mu
\quad, \pi^{-} + \delta \Longrightarrow W^{-} \Longrightarrow \mu^{-} +
\tilde{\nu}_\mu ,
\end{eqnarray}

\vspace{-0.7cm}

\begin{eqnarray}\label{hn}
\pi^{+} + \delta \Longrightarrow W^{+} \Longrightarrow e^{+} + \nu_e
\quad, \pi^{-} + \delta \Longrightarrow W^{-} \Longrightarrow e^{-} +
\tilde{\nu}_e ,
\end{eqnarray}

 For the first time in a hundred-years history of an electron, common known as
the smallest stable ElmMicrPrt, there exist a posibility for a consideration
of its unusual behaviour by means of a transparent PhsMdl realized in a
natural way without any irreconcilable contradictions. I cherish hope that
our consideration from quit new point of view of my PhsMdl of all leptons by
means of the PhsMdl of the electron will be of great interes for all
scientists. Our PhsMdl explain as the structure of leptons and adrons and the
nature of their interaction so the existence of a possibility for joint
description of a field and substantial form of the matter as unity whole in
the physical science, which are submitted to an united, fundamental and
invariable laws of nature.


\begin{thebibliography}{99}

\bibitem{JMRb} Rangelov J.M., University Annual (Technical Physics), {\bf 22},
(2), 65, 87, (1985) ; {\bf 23}, (2), 43, 61, (1986) ; {\bf 24}, (2), 287,
(1986); {\bf 25}, (2), 89, 113, (1988).

\bibitem{JMRc} Rangelov J.M.,Comptens Rendus e l'Academie Bulgarien Sciences,
{\bf 39}, (12), 37, (1986).

\bibitem{JMRd} Rangelov J.M., Report Series of Symposium on the Foundations
of Modern Physics, 6/8, August, 1987, Joensuu,p.95-99, FTL, 131, Turqu,
Finland; Problems in Quantum Physics'2, Gdansk'89,  18-23, September, 1989,
Gdansk, p.461-487,  World Scientific, Singapur, (1990);

\bibitem{JMRe} Rangelov J.M., Abstracts Booklet of 29th Anual Conference of
the University of Peoples' Friendship, Moscow 17-31 May 1993  Physical ser.;
 Abstracts Booklet of Symposium on the Foundations of Modern Physics, 13/16,
 June,(1994),  Helsinki, Finland  60-62.

\bibitem{JMRf} Rangelov J.M., Abstract Booklet of B R U-2, 12-14, September,
 (1994), Ismir, Turkey ; Balk.Phys.Soc.,  {\bf 2}, (2), 1974, (1994).
 Abstract Booklet of B R U -3,  2-5  September,(1997), Cluj-Napoca,Romania.

\bibitem{JMRi} Rangelov J.M., LANL quant-ph 0001078; 0001079; 0001099;
0002018; 0002019.

\bibitem{HFEH} H.Frauenfelder, E.M.Henley, Subatomic Physics , Prentice-Hall,
(1974)

\bibitem{KN} K.Nishijima, Fundamental Particles, W.A.Benjamin inc. New York
(1964)

\bibitem{KGVW} K.Gattfried, V.F.Weisscopf, Concepts of Particle Physics,
Clarendon Press, Oxford, (1984)

\bibitem{GK} Gunnar Kallen, Elementary Particle Physics, Addison-Wesley
publisher, Comp. Massachusetts, London (1966)

\bibitem{LR} Lewis Ryder, Elementary Particles and Symmetries. Gordon and
Breach Science Publisher New York (1975).

\bibitem{GF} G.Feinberg, What is the world made of? Anchor press/Doubleday,
New York (1978)

\bibitem{FEC} F.E.Close, An Introduction to Quarks and Partons, Academic
press, London (1979)

\bibitem{SOSSMN} S.Ogawa, S.Sawada, M.Nakagava, Composite Model of Elementary
particles, Iwanami Shoten, (1980).

\bibitem{KH} Kerson Huang, Quarks, Leptons and Gauge Fields. World Scientific,
New York, (1982).

\bibitem{DVSh} Red.D.V.Shirkov, Small Encyclopeadia of Micro World Physics.
(in Russian) Sovietscaja encyclopeadia, Moscow (1980)

\bibitem{HF} Harald Fritzsch, Quarks, Verlag Munchen  (1983)

\bibitem{FHAM} F.Halzen, A.D.Martin, Quarks and Leptons. John Wilew$\&$Sons,
New York (1984).

\bibitem{LVO} L.V.Okun, Elementary Particle Physics. (in Russian) Nauka,
Moscow (1988)

\end{thebibliography}
\end{document}